# Artificial Neural Network Based Breast Cancer Screening: A Comprehensive Review

**Subrato Bharati[1], Prajoy Podder[2], M. Rubaiyat Hossain Mondal[3]**

Institute of Information and Communication Technology,
Bangladesh University of Engineering and Technology
Dhaka, Bangladesh
[1]*subratobharati1@gmail.com,*[2]*prajoypodder@gmail.com,* [3]*rubaiyat97@yahoo.com*

*Abstract*: Breast cancer is a common fatal disease for women. Early diagnosis and detection is necessary in order to improve the prognosis of breast cancer affected people. For predicting breast cancer, several automated systems are already developed using different medical imaging modalities. This paper provides a systematic review of the literature on artificial neural network (ANN) based models for the diagnosis of breast cancer via mammography. The advantages and limitations of different ANN models including spiking neural network (SNN), deep belief network (DBN), convolutional neural network (CNN), multilayer neural network (MLNN), stacked autoencoders (SAE), and stacked de-noising autoencoders (SDAE) are described in this review. The review also shows that the studies related to breast cancer detection applied different deep learning models to a number of publicly available datasets. For comparing the performance of the models, different metrics such as accuracy, precision, recall, etc. were used in the existing studies. It is found that the best performance was achieved by residual neural network (ResNet)-50 and ResNet-101 models of CNN algorithm.

## I. Introduction

Breast cancer is a very fatal and common disease for women worldwide. There are several cancer types such as liver cancer, breast cancer, lung cancer, brain cancer, and so on where the third common cancer is breast cancer among the several types of cancers. According to the world health organization (WHO), 11.3% (1.7 million) of patients affected in 2015 has been connected to breast cancer (WHO 2018). Moreover, the several quantity of new breast cancer affected patients is estimated to grow by 75% in the next 20 years. As a result, precise and early diagnosis shows a crucial role to develop the diagnosis and rise the patients' survival rate with breast cancer from 20% to 60% according to the WHO in 2019. Generally, there are 2 types of breast tumors such as malignant and benign. Malignant is the cancerous tumor (invasive) and benign is the non-cancerous tumor (noninvasive) [1-2]. Those tumors have more subtypes that require to be detected separately as each can lead to treatment plans and various prognosis. Breast cancer with its subcategories need accurate diagnosis, which is known as multi-classification. Medical imaging systems are more simply effective and adopted for breast and lung cancer recognition than new testing methods [3, 4]. Some of the well-known medical imaging systems for the diagnosis of breast cancer are breast X-ray mammography, sonograms or ultrasound imaging, computed tomography, MRI, and histopathology images [5-7]. Figure 1 shows an example of breast mammography, while Figure 2 shows an example of screening of mammograms. Medical imaging is generally done by manually through one or other skilled doctors (such as sinologists, radiologists, or pathologists). The complete decision is prepared after consent if several pathologists are present for breast cancer histopathology image analysis; otherwise findings are described by a pathologist. However, manual histopathology image investigation have several issues [8, 9]. Firstly, expert pathologists are rare in some low income and developing countries. Secondly, the technique of multi-class classification with image analysis is time consuming and cumbersome for pathologists. Thirdly, pathologists may have deteriorated attention and may experience fatigue at the time of image analysis. Finally, a consistent of subtype identification of breast cancer depends on the domain knowledge and professional experience of a skilled pathologist. Especially, these issues are for the breast cancer early stage and caused misdiagnosis. Conversely, computer-aided diagnosis (CAD) schemes can help as a second outlook to resolve multi-classification problems for breast cancer. A CAD scheme is an inexpensive, fast, reliable source, and readily available of cancer early diagnosis [10, 11]. About 30% to 70% reduction in the morality rate can be achieved in this process [12]. The introduction of digital medical images has given an edge to artificial intelligence (AI) for pattern recognition conducting a CAD scheme. CAD schemes are considered to assist physicians by interpreting automatic images. Therefore, such a system diminishes human dependency, rises diagnosis rate, and decreases the total treatment expenses by decreasing false negative (FN) and false positive (FP) predictions [13]. Furthermore, higher FN rate may cause breast cancer carriers with no treatment, and misdiagnoses occur in the breast cancer early stages. It is described in [11] that a CAD scheme uses for the classification which rises sensitivity to 10%. Despite of classifications [14, 15], CAD systems may be established to implement new diagnosis-related tasks, for example, lesion detection [16, 17], registration, segmentation [18, 19], and grading [20, 21].

Recently, numerous research papers have been proposed to explain breast cancer classification [1], registration, detection, segmentation, and grading problems by conducting machine learning schemes such as, naïve Bayes, random forest, support vector machine (SVM), decision tree, or by conducting artificial neural networks (ANN)-based approaches, deep neural networks (DNN) and spiking neural network (SNN).

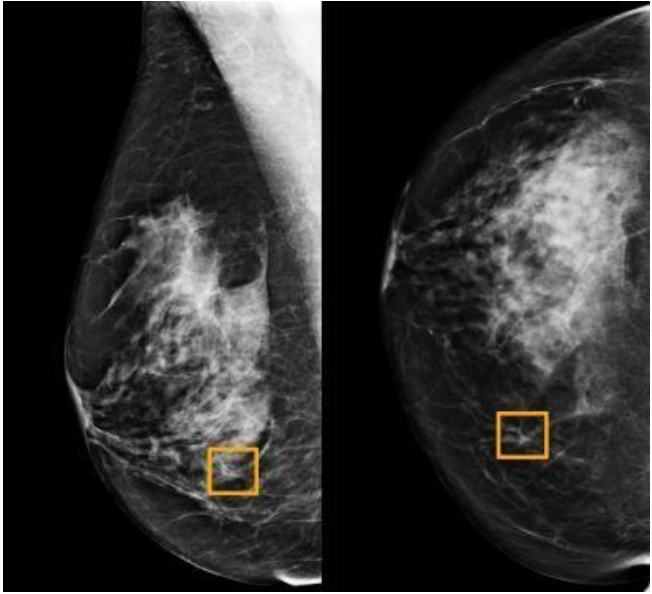

**Figure 1.** Breast Mammography (Adapted from [98])

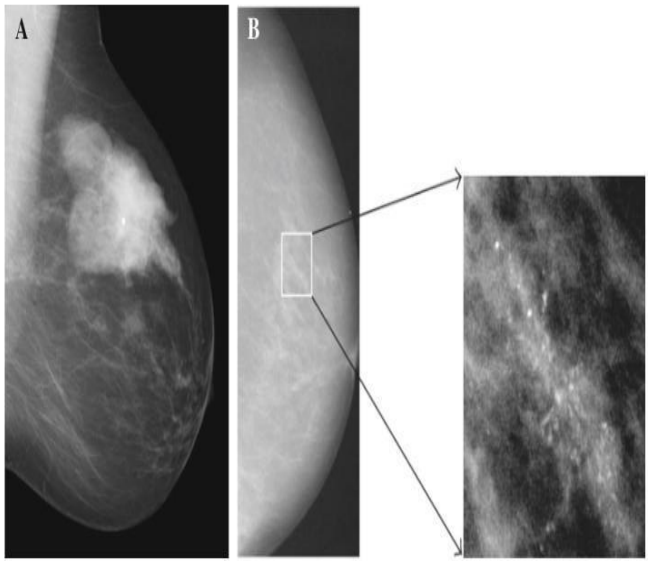

**Figure 2.** Screening of mammograms (a) normal view (b) magnified view with micro-calcifications (Adapted from [86, 97])

This paper provides a comprehensive review of the application of ANN models for mammographic detection of breast cancer. The rest of the paper can be structured as follows. Section II illustrates ANN approaches used for breast cancer classification. Section III describes different computer-aided diagnosis (CAD) systems using deep learning. Section IV illustrates the result analysis of different deep learning models for breast cancer classification. Limitations of the existing research are described in Section V, while the direction for future research is presented in Section VI. Finally, a conclusion is provided in Section VII.

## II. Breast Cancer Classification using ANN

Figure 3 shows a sample of ANN consisting of input, hidden and output layers. An ANN is a machine learning algorithm suitable for different tasks including classification, prediction and visualization. Furthermore, an ANN is suitable or multi-disciplinary tasks with the use of multiple types of data which may be unstructured, semi-structured and structured data. For breast cancer medical images which are a form of unstructured data, shallow ANN and DNN are considered. Table 1 summarizes ANN models for breast cancer diagnosis. Figure 4 and Figure 5 show some categories under DNN and Deep CNN.

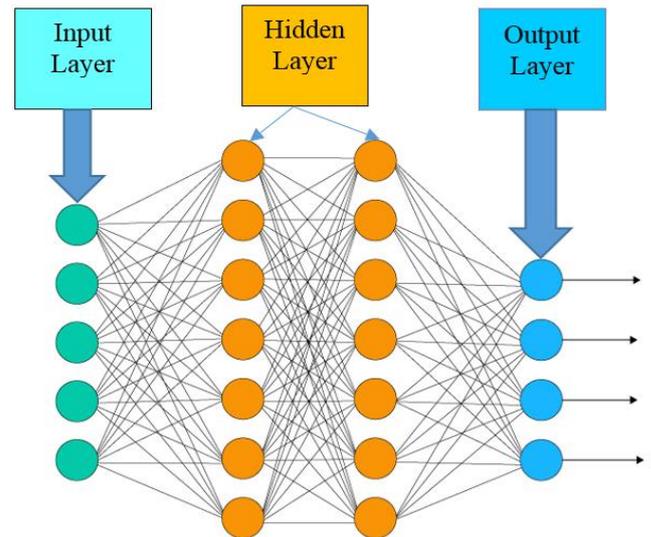

**Figure 3.** Sample of ANN

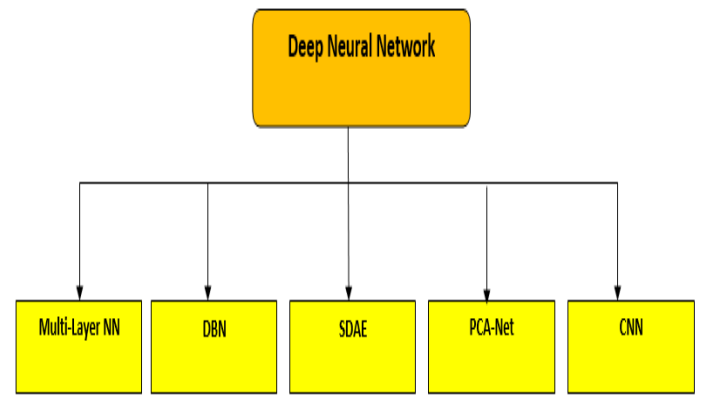

**Figure 4.** DNN Models in breast Cancer

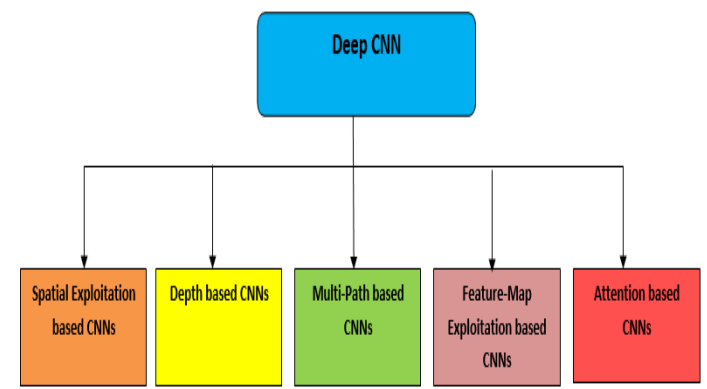

**Figure 5.** Categories of Deep CNN architecture

| **Types of ANN** | **Strengths** | **Weaknesses** |
|---|---|---|
| Machine Learning driven NN | (i) Network size is small.<br>(ii) Less training time is required.<br>(iii) Requires less memory.<br>(iv) Better generalization performance can be achieved using the additional hidden layers. | (i) Cannot provide good results when data is high dimensional.<br>(ii) More data is needed to obtain good performance for the additional hidden layer. |
| SNN | (i) Network size is small.<br>(ii) Less training time is required.<br>(iii) Easy to train<br>(iv) Not complex to optimize the training parameters for achieving better accuracy<br>(v) Small scale data can also achieve good performance. | (i) Cannot provide good results when data is high dimensional.<br>(ii) Performance depends on the structure of ANN and also on the features that are designed.<br>(iii) Not easy to generalize the predictive results. |
| DBN | (i) Automatic de-noising system for high dimensional data improves the breast cancer classification with its performance in real medical images.<br>(ii) It is the models learning like backpropagation that may minimize cross entropy. | (i) The log likelihood is to unable the track the loss. |
| SDAE | (i) This greedy learning with efficiency can be monitored. Combined with other learning techniques that the all weights fine-tune to develop the discriminative or generative performance of the entire network.<br>(ii) It can be used for HD data that keep correlated features. | (i) Breast cancer images can be affected by noise. De-noising process in order to eliminate the undesired noise performs better on image data having high dimension. |
| PCA-Net | (i) Owing to big receptive field, PCANet can extract whole explanations of the objects' images and seizures more information according to semantic level.<br>(ii) Owing to block histogram and binary hashing PCA-Net is very flexible for justification and mathematical analysis of its usefulness. | (i) The procedure of simple hashing scheme has not provided rich sufficient information to draw the features.<br>Therefore, it effects the performance of representation desired when possess of data has numerous irrelevant information. |
| CNN | (i) Performs well to extract the relevant information from the image having fewer weights in its layers.<br>(ii) Automatically detects the desired features without the human supervision.<br>(iii) Computational cost is low.<br>(iv) CNN performs better in disease detection fast compared to others. | (i) Provides poor result for small target datasets such as 80 images.<br>(ii) CNN fine-tuned model (FTM-ARL) needs more training time compared to the other CNN types. Since in this model new layers need to be trained from the scratch. ARL means append or remove layer. |

*Table 1.* ANN model for breast cancer with its summary

## III. CAD Models and its Basic Concept

A CAD model has been conducted to offer extra information and maintain the decision creation on cancer staging and disease diagnosis. It is dissimilar from a CAD model which targets to detect, segment, or localize suspicious regions. Conversely, it has been detected that a CAD model could be located ahead of an identification model for complete investigation from the localization and detection to the suspicious regions diagnosis.

### A. ML-Based CAD

A ML-based CAD system refers to ML-based classification and feature extraction as visualized in Figure 6 where feature selection scheme is elective in this system. Generally conducted features derive from image descriptors which quantify the shape, intensity, and suspicious region textures [22]. Desired ML classifiers have not been limited to SVM [91], ANN, KNN, Random Forest, Naïve Bayesian [23]. Owing to the radiomics emergency, it has been well-known that feature selection can be very significant and it targets to suspicious lesions feature with retrieve intrinsic.
Figure 6 shows the flow diagram of the ML based CAD architecture.

### B. CNN-Based CAD

CNN is a kind of computational model composed of multiple layers to recover features as raw data. It represents hierarchical abstraction [24]. CNN model is visualized in Figure 7.

A DNN consists of a number of layers including convolution layers, fully connected (FC) layers, pooling layers, and an output layer. Among these layers, a convolution layer are useful for learning high-level features for example the edges of an image. FC layers are for learning features at the pixel level. A pooling layer can reduce the size of convolved features resulting in the reduction in required computational power. This layer can perform two types of operations, max pooling and average pooling [87, 88].

The CNN used for breast image or breast data classification can be categorized into two sections, de novo trained model and transfer learning based model. The CNN based models generated and trained from the scratch are denoted as "de novo model" [89]. On the contrary, CNN models utilizing previously trained neural network models such as AlexNet, visual geometry group (VGG), residual neural network (ResNet), etc. are called "transfer learning (TL)-based models" [92].
Figure 8 displays the VGG16 architecture which refers to thirteen convolutional layers, five pooling layers, three full-connection layers, and one softmax layer [25].
For more development in the classification of object, several methods have been embedded, containing nonlinear filtering, normalization of local response, data augmentation, multiscale representation, and hyperparameter optimization [26, 27]. There are several deep learning systems such as AlexNet [28], VGG [29], ResNet [30], LeNet [31], you only look once (YOLO) [32], GoogLeNet [33, 34], LSTM [35], and faster R-CNN [36].
Note that CNN may be trained end on and can be data-driven. It enables the feature extraction with integration, feature selection, as well as malignancy prediction according to an optimization technique. Consequently, human engineers cannot design these retrieved features from the input. Generally, CNN-based CAD provides a remarkable performance which comes from hardware resource in advanced computing (i.e., distributed computing and GPU), open-source software, i.e., TensorFlow, and HQ labeled images with open challenges, i.e., ImageNet. It also gets advantages from the new architectures design of deep learning, i.e., identity and inception mapping.

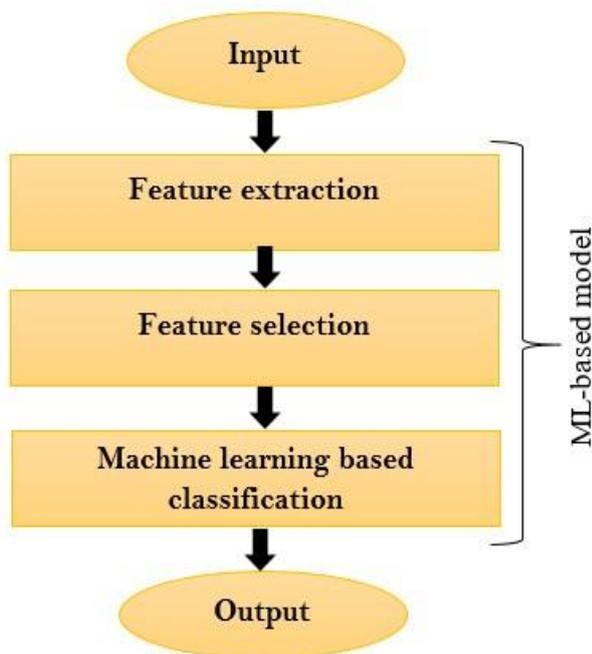

**Figure 6.** ML-based CAD architecture

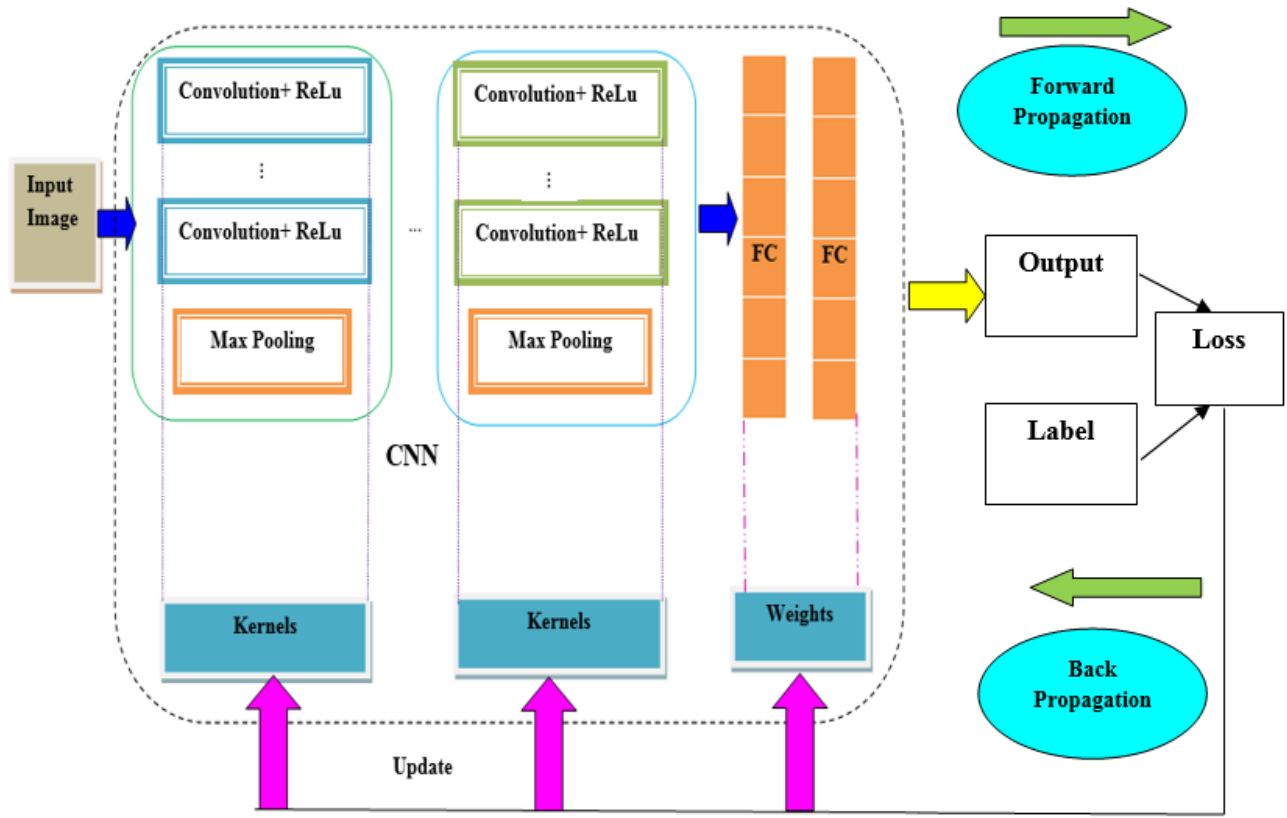

**Figure 7.** Basic CNN-based CAD

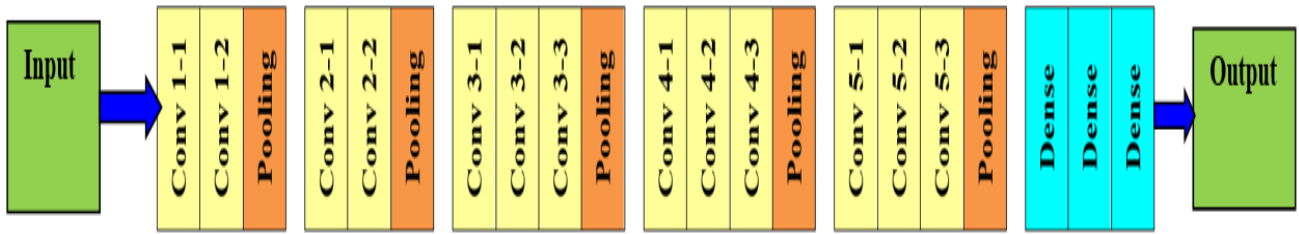

**Figure 8.** Basic VGG-16 model

## IV. Result Analysis of different ANN models

Many studies use single CNN without a fusion. A number of research works [67, 82] embeds some residual blocks in the convolutional layer on the basis of pre trained models. The combination of convolution layer and residual block showed good performance with an accuracy of 92.19% [82]. However, the work in [82] showed that CNN showed much better performance than fusion models such as long short-term memory (LSTM), and a fusion of CNN and LSTM. This result was for a sample of small number of images which might have influenced the finding.

Table 2 presents different DNN models applied to several datasets of breast cancer classification. In this case, most studies perform classification tasks using CNN instead of SNN or multi-layer NN. The datasets used for classification are DDSM, INBreast, BCDR-FO3, and min-MIAS.

The performance of the models are evaluated using confusion matrix where actual classes and predicted classes can be placed in rows and columns, respectively. Thus, breast cancer can be classified as true positive when correctly classified, or classified as false negative when incorrectly classified. Other metrics for evaluation of breast cancer classification are testing accuracy, sensitivity, specificity, precision, area under the receiver operation characteristics curve (AUC), F1-measure, etc.

The performance results of various models adopted by researchers on different types of breast cancer datasets are summarized in Table 2 where the value of achieved accuracy, AUC, sensitivity and specificity are described.

| Publication year | Quantity of images | Database | Model | Accuracy | AUC | Sensitivity | Specificity | Ref. |
|---|---|---|---|---|---|---|---|---|
| 2018 | 736 | BCDR-F03 | GoogLeNet | 81% | 88% | - | - | [37] |
| 2018 | 736 | BCDR-F03 | AlexNet | 83% | 79% | - | - | [37] |
| 2018 | 736 | BCDR-F03 | Shallow CNN | 73% | 82% | - | - | [37] |
| 2018 | 115 | INbreast | Faster R-CNN | - | 95% | - | - | [38] |
| 2018 | 82,000 | DREAM | Faster R-CNN | - | 85% | - | - | [38] |
| 2018 | 600 | DDSM | ROI based CNN | 97% | - | - | - | [39] |
| 2018 | 5316 | DDSM | Inception V3 | 97.35% (±0.80) | 98% | - | - | [40] |
| 2018 | 200 | INbreast | Inception V3 | 95.50% (±2.00) | 97% | - | - | [40] |
| 2018 | 600 | BCDR-F03 | Inception V3 | 96.67% (±0.85) | 96% | - | - | [40] |
| 2018 | 5316 | DDSM | VGG16 | 97.12% (±0.30) | - | - | - | [40] |
| 2018 | 5316 | DDSM | ResNet50 | 97.27% (±0.34) | - | - | - | [40] |
| 2018 | 120 | MIAS | Deep CNN | 96.7% | - | - | - | [41] |
| 2018 | 20,000 | Private dataset (University of Pittsburgh) | AlexNet, Transfer Learning | - | 98.82% | - | - | [54] |
| 2018 | 2620,115, 847 | DDSM INbreast and private dataset by Semmelweis University Budapest | Faster R-CNN | - | 95% | - | - | [42] |
| 2018 | 78 | FFDM | CNN | - | 81% | - | - | [43] |
| 2018 | 736 | BCDR-F03 | MV-DNN | 85.2% | 89.1% | - | - | [44] |
| 2018 | 322 | MIAS | Deep CNN | 65% | - | - | - | [92] |
| 2017 | 3158 | FFDM | Deep CNN | 82% | 88% | 81% | 72% | [45] |
| 2017 | 115 | INbreast | CNN (COM) | 95% | 91% | - | - | [46] |
| 2017 | 2242 (1057 malignant, 1397 benign) | SFM, DM | Deep CNN | - | 82% | - | - | [47] |
| 2017 | 2796 | IRMA | CNN-CT | 83.74% | 83.9% | 79.7% | 85.4% | [48] |
| 2017 | 2796 | IRMA | CNN-WT | 81.83% | 83.9% | 78.2% | 83.3% | [48] |
| 2017 | 245 | FFDM | VGG19 | - | 86% | - | - | [49] |
| 2017 | 560 | FFDM | Custom CNN | - | 79% | - | - | [50] |
| 2017 | 2795 | IRMA | VGG16 | 100% | 100% | | | [64] |
| 2016 | 600 | DDSM | Deep CNN | 96.7% | - | - | - | [51] |
| 2016 | 607 (219 lesions) | FFDM | AlexNet | - | 86% | - | - | [52] |

| Year | Dataset size | Database | Method | Accuracy | Sensitivity | Specificity | AUC | Ref |
|---|---|---|---|---|---|---|---|---|
| 2016 | 736 (426 benign, 310 malignant lesions) | BCDR- F03 | CNN | - | 82% | - | - | [53] |
| 2017 | 480 | DDSM | SNN | 79.5% | - | - | - | [76] |
| 2017 | - | MIAS, CBIS-INBreast | CNN (COM) | 57% | 77% | - | - | [76] |
| 2018 | 58 | ED (HP) | SDAE | 98.27% (Benign), 90.54% (Malignant) | - | 97.92% (Benign), 90.17% (Malignant) | | [77] |
| 2016 | - | UCI, DDSM | SNN | 89.175%, 86% | - | - | - | [78] |
| 2018 | - | BreakHis | CNN (UDM) | 96.15%, 98.33% (2 Classes), 83.31-88.23 % (8 Classes), | - | - | - | [79] |
| 2016 | - | ED(US) | ML-NN | 98.98% | 98% | - | - | [80] |
| 2017 | 1057 malignant, 1397 benign | ED(Mg),DDSM | Multitask DNN | 82% | - | - | - | [81] |
| 2017 | - | ED (HP) | CNN (COM) | 95.9% (2 classes), 96.4% (15 classes), | - | - | - | [82] |
| 2016 | - | ED(US-SWE) | DBN | 93.4% | 94.7% | 88.6% | 97.1% | [83] |
| 2017 | - | BreakHis | ImageNet | 93.2% | - | - | - | [84] |
| 2018 | 400× (× represents magnification factor) | BreakHis | CNN-CH | 96% | - | 97.79% | 90.16% | [85] |
| 2018 | 400× (× represents magnification factor) | BreakHis | CNN-CH | 97.19% | - | 98.20% | 94.94% | [85] |
| 2020 | 2620 | DDSM [90] | InceptionV3 | 79.6% | - | 89.1% | - | [89] |
| 2020 | 2620 | DDSM [90] | ResNet 50 | 85.7% | - | 87.3% | - | [89] |
| 2020 | 10713 | DDSM patch | ResNet50 | 75.1% | - | - | - | [95] |
| 2020 | 10713 | DDSM patch | Mobile Net | 77.2% | - | - | - | [95] |
| 2020 | 10713 | DDSM patch | MVGG16 | 80.8% | - | - | - | [95] |
| 2020 | 10713 | DDSM patch | MVGG16 + ImageNet | 88.3% | 93.3% | - | - | [95] |
| 2019 | 190 | DDSM | CNN | 93.24% | - | 92.91% | 91.92% | [93] |
| 2019 | 190 | DDSM | CNN based LBP | 96.32% | 97% | 96.81% | 95.83% | [93] |
| 2020 | 292 | DDSM | GAN and CNN | 80% | 80% | - | - | [94] |

*Table 2.* Comparison of results

| CNN Model | Scratch Training Scenario | | | | Initialization on pre-trained weights | | | |
| --- | --- | --- | --- | --- | --- | --- | --- | --- |
| | DDSM-400 | | CBIS-DDSM | | DDSM-400 | | CBIS-DDSM | |
| | **Accuracy** | **AUC** | **Accuracy** | **AUC** | **Accuracy** | **AUC** | **Accuracy** | **AUC** |
| AlexNet | 61% | 65.7% | 65.6% | 71.6% | 73.3% | 80.5% | 75.3% | 80.2% |
| ResNet-50 | 54.8% | 59.5% | 62.7% | 63.7% | 74.3% | 85.6% | 74.9% | 80.4% |
| ResNet-101 | 58.8% | 63.7% | 66.2% | 64.1% | 78.5% | 85.9% | 75.3% | 79.1% |
| ResNet-152 | 54.3% | 59.6% | 64.7% | 60.9% | 63% | 78.6% | 75.5% | 79.3% |
| VGG16 | 59% | 62.1% | 58% | 70.2% | 74.8% | 84.4% | 71.6% | 78.1% |
| VGG19 | 58.8% | 64.4% | 58.1% | 70.7% | 73.8% | 83.5% | 73.6% | 78.3% |
| GoogleNet | 56.9% | 58% | 59.8% | 59% | 75.8% | 83% | 72% | 76.7% |
| Inception-v2 | 59% | 65.2% | 65.4% | 57.7% | 78% | 85% | 74.7% | 77.4% |

*Table 3.* Performance of deep neural networks using the from-scratch training scenario

The performance of multiple networks performed in [96] is summarized in Table 3 from scratch training scenarios and fine tuning scenarios, respectively.

| Pre-processing technique | Approach | Ref. |
| --- | --- | --- |
| Augmentation | Geometric Transform | [54-60] |
| | Patch creation Approaches | |
| | Distortion/Add noise | |
| Scaling | Methods like Bi-cubic interpolation, Gaussian Pyramid, Bilinear interpolation | [61-65] |
| ROI Extraction | Methods conducted like Nuclei Segmentation, region growing, Markov Random, Otsu Method | [55, 66-67] |
| Remove Artifacts | Using the pixel intensity thresholding, and binary images, Extracting Bigger areas, cropping border | [20, 58, 61, 68] |
| Enhancement and Normalization | Adaptive Mean, Histogram equalization, Log transforms, Median filters, Wiener Filter, CLAHE technique | [62,69-72] |
| Stain Normalization or Removal | Color Deconvolution, Stain Normalization | [62,73-75] |

*Table 4.* Studies of pre-processing techniques

AUC and classification accuracy values were used as performance metrics in [96]. They achieved maximum performance using fine-tuning in ResNet-50 and ResNet-101 models in both datasets.

Table 4 illustrates some pre-processing techniques that researchers used in their papers.

## V. Limitations

In the literature, most of the models used publicly available free datasets for BrC classification. This is because the deep learning methods used for BrC classification actually requires a large number of annotated images which is difficult to obtain. A large dataset of breast cancer images require many images and require expert doctors to label the images perfectly. The creation of datasets is thus time consuming and difficult. Hence, researchers prefer to use existing publicly available datasets. However, these models applied to public datasets may not be the optimum solution. These models may be less effective when applied to real-life cancer images.

According to the literature review, SNN and DNN are used for classification of BrC. Some studies preferred SNN as SNN performs better than DNN for small datasets. However, SNN does not work well for high dimensional data and BrC datasets can be high dimensional. On the other hand, DNN based on ML-NN or CNN work well for high dimensional multicast BrC image datasets. Furthermore, the previous studies used two types of CNN which are de novo model (trained from scratch) and pre-trained TL-based model.

## VI. Direction for Future Research

This section represents novel research direction in order to exploit the classification of breast cancer. One direction for classification of breast cancer is the reinforcement learning method supporting ANN. The system architecture and parameter tuning of its environment are the major challenges in recent time. The novel prospects of ML is hybrid ML or ANN technique. Case-based reasoning in breast cancer classification is a method to solve novel problems by the solution of previous problem. It updates and retains the present solution prepared through humans and solves future problems.

## VII. Conclusion

This study represents a review for breast cancer mammographic screening where we describe CAD systems and ANN. We have focused the strengths and weaknesses of some ANN models such as ML-NN, SNN, SADE, DBN, and PCA-Net. Moreover, after the basic concept of CNN-based CAD, we have also provided CNN-based system evaluation where we have included ACC, AUC, SEN, and SPE scores. Furthermore, we have described data pre-processing systems with these approaches. For in-depth evaluation, ANN types (such as SNN, DBN, CNN etc.) are described for different databases. Finally, we have outlined some future scope, challenges, and limitations of this research field.

## Author Biographies

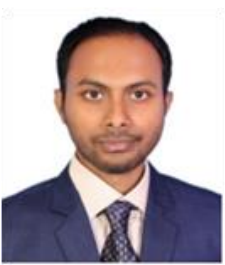
**Subrato Bharati** received his B.S degree in Electrical and Electronic Engineering from Ranada Prasad Shaha University, Narayanganj-1400, Bangladesh. He is currently working as a research assistant at the Institute of Information and Communication Technology, Bangladesh University of Engineering and Technology, Dhaka, Bangladesh. He is a regular reviewer of ISA Transactions, Elsevier; Array, Elsevier; Vehicular Communications, Elsevier; Journal of Systems Architecture, Elsevier; Cognitive Systems Research, Elsevier; Soft Computing, Springer; Data in Brief, Elsevier, Wireless Personal Communications, Springer; Informatics in Medicine Unlocked, Elsevier. He is the guest editor of Special Issue on Development of Advanced Wireless Communications, Networks and Sensors in American Journal of Networks and Communications. His research interest includes bioinformatics, medical image processing, pattern recognition, deep learning, wireless communications, data analytics, machine learning, neural networks, distributed sensor networks, parallel and distributed computing computer networking, digital signal processing, telecommunication and feature selection. He published several IEEE, Springer reputed conference papers and also published several journals paper, Springer Book chapters.

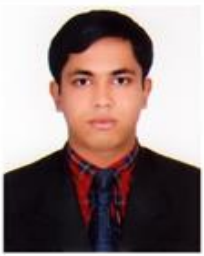
**Prajoy Podder** worked as a Lecturer in the Department of Electrical and Electronic Engineering in Ranada Prasad Shaha University, Narayanganj-1400, Bangladesh. He completed B.Sc. (Engg.) degree in Electronics and Communication Engineering from Khulna University of Engineering & Technology, Khulna-9203, Bangladesh. He is currently pursuing M.Sc. (Engg.) degree in Institute of Information and Communication Technology from Bangladesh University of Engineering and Technology, Dhaka-1000, Bangladesh. He is a researcher in the Institute of Information and Communication Technology, Bangladesh University of Engineering & Technology, Dhaka-1000, Bangladesh. He is regular reviewer of Data in Brief, Elsevier and Frontiers of Information Technology and Electronic Engineering, Springer, ARRAY, Elsevier. He is the lead guest editor of Special Issue on Development of Advanced Wireless Communications, Networks and Sensors in American Journal of Networks and Communications. His research interest includes machine learning, pattern recognition, neural networks, computer networking, distributed sensor networks, parallel and distributed computing, VLSI system design, image processing, embedded system design, data analytics. He published several IEEE conference papers, journals and Springer Book Chapters.

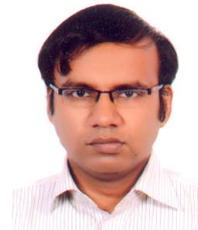
**M. Rubaiyat Hossain Mondal**, PhD is currently working as an Associate Professor in the Institute of Information and Communication Technology (IICT) at Bangladesh University of Engineering and Technology (BUET), Bangladesh. He received his Bachelor's degree and Master's degree in Electrical and Electronic Engineering from BUET. He joined IICT, BUET as a faculty member in 2005. From 2010 to 2014 he was with the Department of Electrical and Computer Systems Engineering (ECSE) of Monash University, Australia from where he obtained his PhD in 2014. He has authored a number of articles in reputed journals published by IEEE, Elsevier, De Gruyter, IET, Springer, PLOS, Wiley and MDPI publishers. He has also authored a number of IEEE conference papers including GLOBECOM 2010 in USA, and presented papers in IEEE conferences in Australia, South Korea, and Bangladesh. In addition, he has coauthored a number of book chapters of reputed publishers which are now in press. He is an active reviewer of several journals published by IEEE, Elsevier and Springer. He was a member of the Technical Committee of different IEEE R10 International conferences. His research interest includes artificial intelligence, bioinformatics, image processing, wireless communication and optical wireless communication.